\documentclass[fleqn,usenatbib,useAMS]{mnras}

\usepackage{graphicx}	
\usepackage{amsmath}	
\usepackage{amssymb}	
\usepackage{multicol}        
\usepackage{bm}		
\usepackage{pdflscape}	

\usepackage[T1]{fontenc}
\usepackage{ae,aecompl}

\usepackage{newtxtext,newtxmath}

\usepackage[normalem]{ulem}

\newcommand{\um}{\,$\mu {\rm m}$\ }

\def\eq#1{{Eq.~(\ref{#1})}}

\title[OIII intensity mapping]{Intensity mapping from the sky: synergizing the joint potential of [OIII] and [CII] surveys at reionization}

\author[]{Hamsa Padmanabhan$^{1, 2}$\thanks{hamsa.padmanabhan@unige.ch}, Patrick  Breysse$^{3,2}$, Adam Lidz$^{4}$ and Eric R. Switzer$^{5}$
\\
$^{1}$D\'epartement de Physique Th\'eorique, Universit\'e de Gen\`eve,
24 quai Ernest-Ansermet, CH 1211 Gen\`eve 4, Switzerland\\
$^{2}$Canadian Institute for Theoretical Astrophysics, 60 St. George Street, Toronto, ON M5S 3 H8, Canada\\
$^{3}$ New York University, New York, NY, USA\\
$^{4}$ University of Pennsylvania, Department of Physics \& Astronomy, 209 S. 33rd Street, Philadelphia, PA 19104, USA\\
$^{5}$ NASA Goddard Space Flight Center, Greenbelt, MD, USA
}

\date{\today}

\pubyear{2021}

\begin{document}
\label{firstpage}
\pagerange{\pageref{firstpage}--\pageref{lastpage}}
\maketitle

\begin{abstract}
We forecast the ability of future-generation experiments to detect the fine-structure lines of the carbon and oxygen ions, [CII] and [OIII] in intensity mapping (IM) from the Epoch of Reionization ($z \sim 6-8$). Combining the latest empirically derived constraints relating the luminosity of the [OIII] line to the ambient star-formation rate, and using them in conjunction with previously derived estimates for the abundance of [CII] in haloes, we predict the expected auto-correlation IM signal to be observed {using new experiments based on the} Fred Young Submillimetre Telescope (FYST) and the balloon-borne facility, Experiment for Cryogenic Large-Aperture Intensity Mapping (EXCLAIM) over $z \sim 5.3 - 7$. We describe how improvements to both the ground-based and balloon-based surveys in the future will enable a cross-correlation signal to be detected at  $\sim$ 10-{30} $\sigma$ over $z \sim 5.3 - 7$. Finally, we propose a space-based mission targeting the [OIII] 88 and 52 $\mu$m lines along with the [CII] 158 $\mu$m line, configured to enhance the signal-to-noise ratio of cross-correlation measurements. We find that such a configuration can achieve a high-significance detection (hundreds of $\sigma$) in both auto- and cross-correlation modes. 
\end{abstract}


\begingroup
\let\clearpage\relax
\endgroup
\newpage

\section{Introduction}

Intensity mapping (IM) is a promising way to recover large-scale cosmological constraints from the line emission of tracers of cosmic structures \citep[for a review, see, e.g.][]{Kovetz:2017agg}, which can potentially access a much larger number of modes than traditional galaxy surveys \citep{loeb2013}. Today, this technique has already been shown to provide constraints on the abundance and clustering of emitting gas in the late-time universe \citep{chang10, masui13, switzer13, 2018MNRAS.476.3382A, 2018MNRAS.481.1320C, 2019MNRAS.489L..53Y, 2016ApJ...830...34K, uzgil2019, 2020ApJ...901..141K} and promises an exciting outlook for the future, including investigations into fundamental physics beyond the standard model \citep[e.g.,][]{hall2013, Fonseca2017, bauer2021, camera2020, bernal2021, liu2020} as well as the physics of galaxy formation \citep[e.g.,][]{wyithe2008b,wolz2016,  hploeb2020, yang2021}. 

Particularly in the sub-millimetre regime (gigahertz and terahertz frequencies), IM of molecular and atomic lines holds promise for studying the evolution of the star-formation rate (SFR) at high-redshifts, and the contribution of molecular gas to this process. Two salient lines of interest to IM in this regime are the [CII] 158\um fine-structure line \citep[e.g.,][]{silva2015, LidzTaylor16, fonscea2017, yue2015, dumitru2018, sun2018, hpcii2019},  and the [OIII] 88\um line, which primarily trace star-forming regions, as the brightest and most commonly observed lines in star-forming galaxies at $z > 6$. Of these, the latter species, [OIII], comes from the second ionization of oxygen, which requires a fairly hard ($> 35.1 \ \text{eV}$) ionization energy, commonly produced by hot O-type stars.  The [OIII] line is extinction-free and sensitive to the electron density, ionization parameter, and gas-phase oxygen abundance of the ionized medium. Examples of galaxies which have been discovered in [OIII] emission near the epoch of reionization include the SPT0311-58E at $z = 6.9$, the Lyman-Break Galaxies A2744\_YD4 at $z = 8.38$ \citep{laporte2017}, the J1211-0118, J0235-0532, and J0217-0208 at $z \sim 6$ \citep{harikane2020}, MACS0416\_Y1 at $z = 8.31$ \citep{tamura2019},  and  MACS1149-JD1 \citep{hashimoto2018} at $z \sim 9.11$. The latter two galaxies show a surprising lack of [CII] \citep{laporte2019}, while SPT0311-58E is detected in [CII] as well.
High values of the [OIII]/[CII] ratio in many of these systems, sometimes labelled the [CII] `deficit' \citep{laporte2019, carniani2017, hashimoto2019}, are associated with a high ionization ratio \citep[e.g.,][]{katz2017}. This is also indicated by the non-detection of [CII] in some systems, although [OIII] is detected \citep[though see also][]{hashimoto2018a, walter2018,inoue2016}. Recently, cosmological simulations of galaxy formation have been used to study the properties of bright [OIII] emitters at $z \sim 9$, finding them to be hosted in dark matter haloes with masses $> 10^{11} M_{\odot}$ \citep{moriwaki2018}.

The above findings add support to a strong science case for detecting [OIII] 88 $\mu$m from wide-field intensity mapping surveys, if its line strength is actually brighter than that of the [CII] line. Perhaps most importantly, another tracer of the large-scale structure (in addition to [CII] and HI 21 cm) may help in alleviating the interloper challenges in [CII] surveys (see e.g. \citealt{Kovetz:2017agg}). Cross-correlations with the HI 21-cm line during the epoch of reionization (EoR)  \citep{visbal2010, lidz2011,gong2012,dumitru2018} are also an important motivation for focusing on [OIII] from EoR galaxies, given that as many consistency tests as possible will strengthen the reliability of the results, especially in view of the challenges such as foregrounds facing current 21-cm observations. Furthermore, measuring [CII] and [OIII] together provides a more complete picture of the ISM than either of the tracers taken separately, especially because [OIII] is associated with the HII regions while [CII] at high redshift originates primarily from photon dominated regions \citep[PDRs;][]{vallini2016, lagache2018}.There are also prospects for measuring the [OIII] 52 \um line with ALMA and JWST \citep{yang2021, jones2020} which together will help break important degeneracies in physical processes governing galaxy formation at these epochs.

\begin{figure}
    \centering
    \includegraphics[width = \columnwidth]{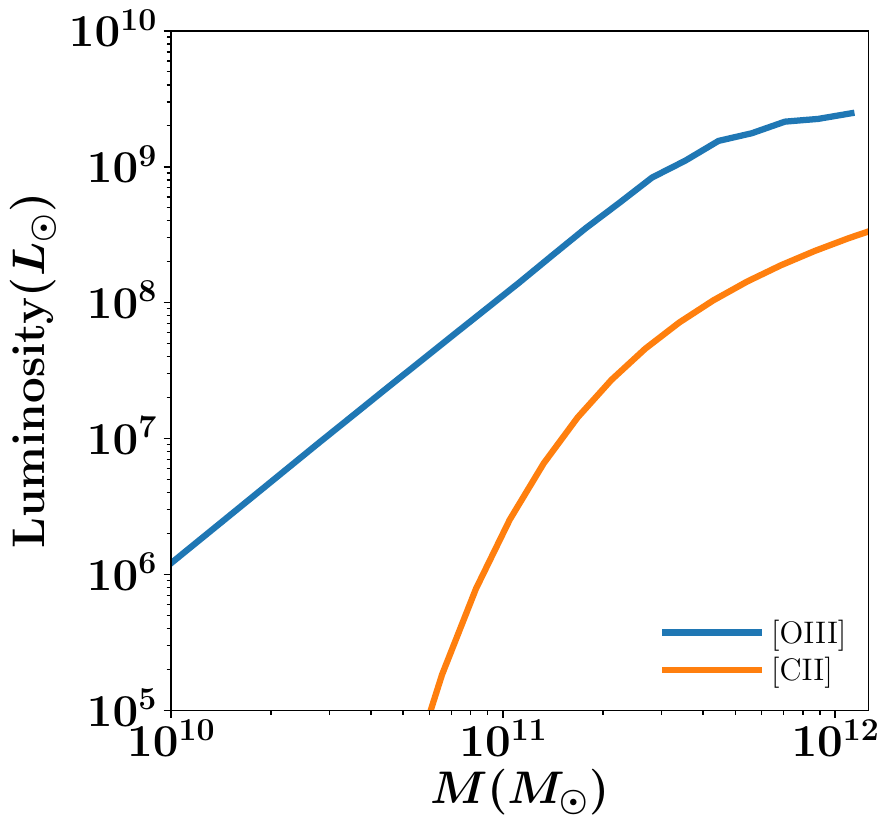}
    \caption{[OIII] and [CII] luminosity - halo mass relations at $z \sim 5.3$, derived using the formalism in the main text. Note that star-forming galaxies are extremely rare in haloes above $10^{12} M_{\odot}$ at $z > 5$, restricting the range of halo masses probed in the plot.}
    \label{fig:oiiilumfunc}
\end{figure} 
In this paper, we detail the prospects for IM with the redshifted [CII] 158\um and [OIII]  lines at high redshifts ($z > 5$) and their synergies with one another. We consider configurations improving upon { the design of} two upcoming experiments in the gigahertz frequency regime: (i) the FYST \citep[the Fred Young Submillimetre Telescope;][]{terry2019}\footnote{https://www.ccatobservatory.org/index.cfm} located at the Cerro Chajnantor in Chile, which covers the frequency range 212 - 428 GHz over an $\sim 8$ sq.deg area and is sensitive to the [CII] 158\um (rest-frame 1900\,GHz transition) in emission in the $5 < z < 9$ redshift range, and 
the EXCLAIM \citep[EXperiment for Cryogenic Large-Aperture Intensity Mapping;][]{exclaimpaper2020}, a high-altitude, balloon spectrometer mission over 420-540 GHz, sensitive to [CII] 158\um in emission over $0 < z < 3.5$, and over a ${\sim}400$ sq. deg area overlapping with the Baryon Oscillation Spectroscopic Survey (BOSS). We outline prospects for the cross-correlation of [CII] and [OIII] using enhanced versions of these two surveys, envisioning plausible future upgrades to each of FYST and EXCLAIM. Finally, we consider a space-based, near-ideal, or `designed' IM survey targeting both the [OIII] 52 \um and 88 \um lines along with the [CII] 158 \um line and whose parameters are tailored to combine the benefits of both the { improved} FYST and EXCLAIM values, and forecast its expected sensitivity. 

The paper is organized as follows. In the following section (Sec. \ref{sec:formalism}), we outline the formalism for deriving the intensity mapping power spectrum of the [CII] and [OIII] at various redshifts. Subsequently (Sec. \ref{sec:autocorr}), we forecast autocorrelation power spectra and signal-to-noise ratios for an improved version of the EXCLAIM-like configuration, and a ground based [CII] survey conducted with an enhanced FYST-like configuration covering the same redshifts, $z \sim 5.3, 6$ and 7. We then (Sec. \ref{sec:crosscorr}) describe the cross-correlation of the above [OIII] and [CII] surveys. Finally, we consider a `design' experimental configuration in Sec. \ref{sec:design}, describing a futuristic space-based mission, in which the parameters are tailored to optimize the advantages of both these configurations, targeting both the [OIII] 88 $\mu$m and 52 $\mu$m lines, in addition to [CII]. We predict the SNR achieved with this configuration in both auto- and cross-correlation modes over $z \sim 5.3, 6$ and 7. We summarize our results and discuss future prospects in a brief concluding section.

\section{Formalism}
\label{sec:formalism}

We begin with the formalism for deriving the intensity mapping power spectrum of the tracers, in this case [CII] and [OIII], given by:

\begin{equation}
 I_{\nu_{\rm obs}, {i}} = \frac{c}{4 \pi} \int_0^{\infty} dz' 
\frac{\epsilon[\nu_{\rm obs, i} (1 + z'), z']}{H(z') (1 + z')^4}
\end{equation} 
In the above expression, the emissivity, $\epsilon(\nu, z)$ is given by:
\begin{equation}
 \epsilon(\nu, z) = \delta_{\rm D}(\nu - \nu_{i}) (1 + z)^3 \int_{M_{\rm min, 
i}}^{\infty} dM \frac{dn}{dM} L_{i}(M,z)
\end{equation} 
The quantity $L_{i}$ represents the luminosity of the tracer $i \in \{\rm [CII], [OIII]\}$ emitting at corresponding frequencies $\nu_i$ and observed at $\nu_{\rm obs, i}$, and as a 
function of their halo mass $M$ and emission redshift $z$. The $M_{\rm min, i}$ denotes the minimum halo mass associated with the respective [CII]/[OIII]-emitting galaxies, with $H(z)$ being the Hubble parameter at redshift $z$.
Using this, the intensity of emission becomes:
\begin{equation}
I_{\nu_{\rm obs}, i} = \frac{c}{4 \pi} \frac{1}{\nu_{i} H(z)}  
\int_{M_{\rm min, i}}^{\infty} dM \frac{dn}{dM} L_{i}(M,z)
\label{CIIspint}
\end{equation} 
if the redshift $z$ is assumed to be the same for both tracers. Here, we use the value of $M_{\rm min} = 10^9 h^{-1} M_{\odot}$ for both [OIII] and [CII] \citep[e.g.,][]{munoz2011, chung2020}.

From this, we can calculate the power spectrum for each of the tracers,  with:
\begin{equation}
 P_{\rm shot, i}(z) = \frac{\int_{M_{\rm min,i}}^{\infty} dM (dn/dM) L_{i} 
(M,z)^2}{\left(\int_{M_{\rm min, i}}^{\infty} dM (dn/dM) L_{i} 
(M,z)\right)^2}
\end{equation}
representing the shot noise component, and 
\begin{equation}
 b_{i}(z) = \frac{\int_{M_{\rm min, i}}^{\infty} dM (dn/dM) L_{i} 
(M,z) b(M,z)}{\int_{M_{\rm min, i}}^{\infty} dM (dn/dM) L_{i} (M,z)}
\end{equation} 
representing the clustering bias of the tracers. Here, $b(M,z)$ denotes the dark 
matter halo bias following \citet{scoccimarro2001}. The full power spectrum becomes:
\begin{equation}
 P_{i}(k,z) =  I_{\nu, i} (z)^2 [b_{i}(z)^2 P_{\rm lin}(k,z) + 
P_{\rm shot, i}(z)]
\label{powerspeccii}
\end{equation} 
with $P_{\rm lin}(k,z)$ being the linear matter power spectrum.

We employ the [CII] luminosity relation from \citet{hpcii2019}:
\begin{equation}
L_{\rm CII}(M,z) = \left(\frac{M}{M_1}\right)^{\beta} \exp(-N_1/M) 
\left(\frac{(1+z)^{2.7}}{1 + [(1+z)/2.9)]^{5.6}} \right)^{\alpha}
\label{lciihighz}
\end{equation}
with the following best-fitting and error values for the free parameters:
\begin{eqnarray}
M_1 &=& (2.39 \pm 1.86) \times 10^{-5}; \  N_1 = (4.19 \pm 3.27) \times 10^{11} M_{\odot}; 
\nonumber \\
 & & \beta = 0.49 \pm 0.38, \ \alpha = 1.79 \pm 0.30
\end{eqnarray}
For [OIII], we follow the best-fitting functional form from the observations of \citet{harikane2020} \footnote{This is also consistent with the findings of \citet{arata2020} from a sample of $z \sim 6-9$ [OIII] emitters.}:
\begin{equation}
\log\left(\frac{L_{\rm OIII}}{L_{\odot}}\right) = 0.97 
\times \log \frac{\rm SFR}{[M_{\odot} \text{yr}^{-1}]} + 7.4
\end{equation}
which connects the [OIII] luminosity to the star formation rate  (SFR) in $z \sim 6-9$ galaxies. We use the empirically derived SFR - halo mass relation ${\rm{SFR}}(M,z)$,  from \citet{behroozi2013} in the above equation\footnote{The \citet{behroozi2013} relation is calibrated through abundance matching to star formation rate data over $z \sim 0-10$, and on halo masses in the range $10^{11} - 10^{13} M_{\odot}$ at $z > 5$.} to infer the $L_{\rm OIII}(M,z)$, assuming no significant evolution of the relation when extrapolated to redshifts below 6. 
A plot of the mean $L_{\rm OIII} - M$ relation so derived\footnote{We neglect scatter in the luminosity-halo mass relation, which is likely to underestimate the shot-noise contributions to the auto-power spectra in each line and in the cross-power if the shot noise in the two lines is correlated.}  is shown in Fig. \ref{fig:oiiilumfunc} along with the corresponding [CII] luminosity - halo mass relation, both at $z \sim 5.3$.

\section{Autocorrelation forecasts for [CII] and [OIII]}
\label{sec:autocorr}

We now forecast the expected auto-correlation signal and noise power spectra in [CII] and [OIII] for FYST-like and EXCLAIM-like surveys covering redshifts $z > 5$.  We explore the prospects for next generation measurements here since we find that autocorrelation in [OIII] or a cross-correlation detection is out of reach for these upcoming surveys, but may be enabled by plausible upgrades to each configuration. 
{We consider both a `Stage II' concept and a more advanced configuration for the [CII] experiment.
For a Stage II concept, we use the noise-effective intensity (NEI) of the base FYST configuration \citep{ccat2021}, rescaled to the per-channel bandwidth used here and also adjusted for a time-multiplexing factor. \footnote{ Specifically, the reported FYST sensitivity in the \citet{ccat2021} configuration is per spectral channel, and needs to be time multiplexed by a factor of 42 to get the data cube NEI. For the next-generation estimate here, we assume that there are 42 parallel versions of the configuration in \citet{ccat2021}, so that the multiplex factor can be neglected (Dongwoo Chung, private communication).}
We also consider a more advanced, Stage III/IV configuration following the specifications in \cite{silva2015,breysse2019} which leads to 2-3 orders of magnitude more improvement in sensitivity.}

\begin{figure}
    \centering
    \includegraphics[width = 1\columnwidth]{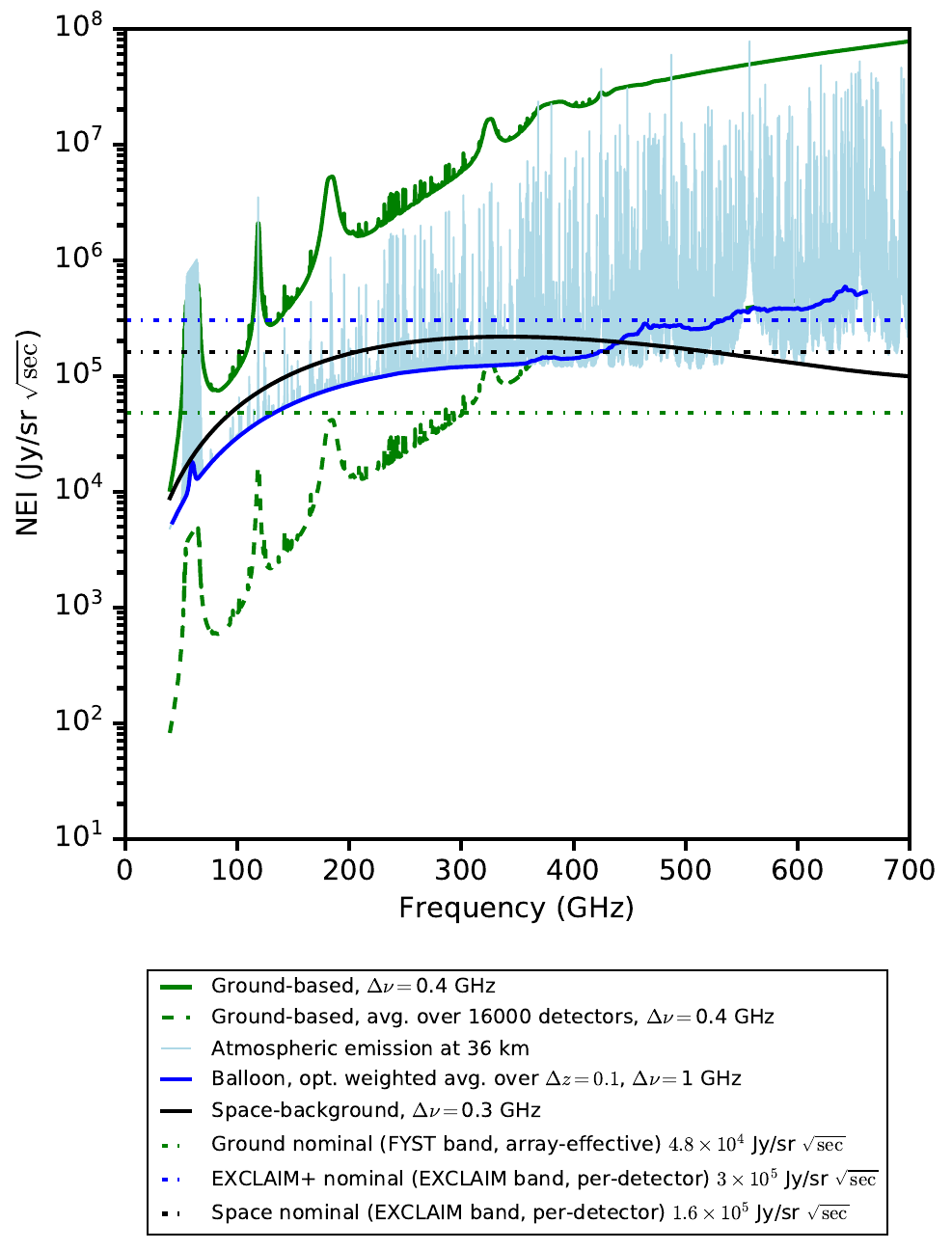} \caption{ Noise-equivalent intensity (NEI) for ground (green), balloon (blue), and space-based (black) platforms. In the ground-based case, we show an array-effective NEI assuming 16000 detectors (dashed green) to average over higher photon loading. For the balloon and space case, we report the per-detector sensitivity. At balloon float altitudes, the Earth's atmosphere resolves into narrow lines due to diminished pressure broadening (light blue). Taking a noise-inverse weighted average of channels (solid dark blue line) allows performance near the darkest windows in the atmosphere. For a next-generation EXCLAIM+ balloon, we assume that the background limit is achieved at $36$\,km float altitude with a focal plane of 30 spectrometers. { This} represents a realistic limit to that instrument approach, yielding ${\approx}5\times$ improved sensitivity over the current EXCLAIM mission \citep{10.1117/1.JATIS.7.4.044004}. Each case (ground, balloon, space) employs different resolution assumptions consistent with the instrument scenarios in Tables\,\ref{table:improved} and\,\ref{table:designexpt}. This results in the space case having higher noise per spectral channel than the balloon platform in some frequencies because it has narrower channels, for example. Horizontal lines show reference noise levels in Tables\,\ref{table:improved} and\,\ref{table:designexpt} assuming ground-based measurements provide the [CII] signal, balloons provide [OIII], and space supplies both [CII] and [OIII] over a wavelength range where the sensitivity varies slowly.}
    \label{fig:nei_scenarios}
\end{figure}

For the balloon-based survey, we assume a $\sigma_{\rm N} = 3 \times 10^5 {\rm \,Jy\, s^{1/2} \, sr^{-1}}$ { (the lowest number that can reasonably be expected in this configuration, see Fig. \ref{fig:nei_scenarios} and following discussion)}.  The parameters of all the above `improved' survey configurations are listed in Table \ref{table:improved}.  
Note that the parameter $\sigma_{\rm N}$ depends on frequency, and we assume in our forecasts in the forthcoming sections that the value of $\sigma_{\rm N}$ for the FYST-like survey is averaged over the band to its value at $z \sim 6$ \citep[e.g.,][]{chung2020}.  
The noise in an autocorrelation survey with the FYST-like or EXCLAIM-like configurations is given by:
\begin{equation}
P_{\rm N, ij} = V_{\rm pix, ij} \frac{\sigma_{\rm N, j}^2}{t_{\rm pix, ij}}
\label{noiseciiauto}
\end{equation}
with the parameter $\sigma_{\rm N, j}$ being the noise-equivalent intensity (NEI) of the experiment $j \in \{\rm EXCLAIM, \ FYST \ Stage \ II, \ FYST \ Stage \ III/IV\}$ as defined in Table \ref{table:improved} and $i \in \{\rm [CII], [OIII]\}$. 

In the above expression, $t_{\rm pix, ij}$ is given by
\begin{equation}
t_{\rm pix, ij} = t_{\rm obs, j} N_{\rm spec, eff, j} \frac{\Omega_{\rm beam, ij}}{S_{A, j}}
\label{eqntpixauto}
\end{equation}
where $t_{\rm obs, j}$ is the total observation time of experiment $j$, and $N_{\rm spec, eff, j}$ is the effective number of pixels of the experimental configuration \citep[e.g.,][]{hpcii2019}. {The latter term is used to quantify the effective number of detectors which integrate in parallel on voxels with a given frequency. For the balloon-based survey, we set $N_{\rm spec, eff}  = 30$}. The  $S_{A, j}$ is the survey area on the sky, and $\Omega_{\rm beam, ij} = 2 \pi (\sigma_{\rm beam, ij})^2 = 2 \pi (\theta_{\rm beam, ij}/\sqrt{8 \ln 2})^2$, with $\theta_{\rm beam, ij} = \lambda_i (1+z)/D_{\rm dish, j}$ being the full width at half maximum of the respective instrument (here, $\lambda_i$ is the rest wavelength of the respective transition, $i \in \{\rm [CII], [OIII]\}$.) 

In both expressions above, the pixel volume, $V_{\rm pix, ij}$ is given by the expression \citep{dumitru2018}:
\begin{eqnarray}
V_{\rm pix, ij} &=& 1.1 \times 10^3  {\rm (cMpc}/h)^3 \left (\frac{\lambda_i}{158  \ 
\mu m} \right) \left(\frac{1 +z}{8} \right)^{1/2}  \nonumber \\
&& \left(\frac{\theta_{\rm beam, j}}{10 '} \right)^2 \left(\frac{\Delta \nu_j}{400 
{\rm MHz}} \right)
\label{eqnvpix}
\end{eqnarray}
where $\Delta \nu_j$ is the spectral resolution of the respective instrument. 
Note that the beam FWHM $\theta_{\rm beam}$ cancels between the pixel volume $V_{\rm pix, ij}$ and the observing time per pixel $t_{\rm pix, ij}$, showing that the noise term $P_{\rm N, ij}$ in \eq{noiseciiauto} is independent of the pixel size used in the configuration.

\begin{table*}
\begin{center}
    \begin{tabular}{ | c | c | c | c | c | c | c | c | c |  p{5 cm} |}
    \hline
    Configuration &  $D_{\rm dish}$ (m.) & $\Delta \nu$ (MHz) &  $N_{\rm spec, 
eff} $ & $S_{\rm A}$ (sq. deg.) & $\sigma_{\rm N}$ 
(Jy s$^{1/2}$ / sr) & $B_{\nu}$ & $t_{\rm obs}$ (h.)  & Survey Bandwidth \\ \hline
   
    EXCLAIM-like & 0.74 & 1000 & 30 & 100 & $3 \times 10^5$  & 40 GHz  & 72 h & 420 - 540 GHz\\
   FYST-like (Stage II)  &  9 & 400 &  1 & 100 & $4.84 \times 10^4$  & 40 GHz & 2000 h & 212 - 428 GHz \\ 
   FYST-like (Stage III/IV)  &  9 & 400 &  16000 & 100 & $2.1 \times 10^5$  & 40 GHz & 2000 h & 212 - 428 GHz \\
\hline
    \end{tabular}
\end{center}
\caption{ {Experimental parameters for improved versions of the EXCLAIM-like and FYST-like survey configurations, which use a lower noise and a higher $N_{\rm spec,eff}$ for the balloon experiment and both a Stage II and a Stage III/IV configuration for the FYST-like survey motivated as described in the main text. The notations are as defined in the main text, and the Survey Bandwidth here refers to the total frequency range assumed to be probed by each configuration.}}
 \label{table:improved}
\end{table*}

{Fig.\,\ref{fig:nei_scenarios} shows the sensitivity to intensity for ground, balloon and space-based detectors with background-limited performance. The loading is based on emission terms from the atmosphere \citep{2019zndo...3406483P}, CMB and Far-IR monopole backgrounds \citep{1998ApJ...508..123F}, galactic cirrus \citep{2011MNRAS.412.1151B}, and Zodiacal light \citep{2002ApJ...578.1009F}. Of these, the atmosphere and CMB dominate over the wavelength range studied here. The conversion to absorbed power employs $dP/dI = \lambda^2 (\Delta \nu) \epsilon$ for single-moded reception across bandwidth $\Delta \nu$ with $\epsilon = 30\%$ nominal efficiency across all cases. For absorbed power $P$, the squared Noise-Effective Power from photons \citep{2003ApOpt424989Z} is ${\rm NEP}^2 = 2 h \nu P + 2 P^2 / \Delta \nu$ (referring to power at the detector), the sum of shot (Poisson) and wave-limited photon noise respectively. Shot noise dominates throughout, but wave noise contributes long-wave of millimeter wavelengths.  Then the noise-equivalent intensity (NEI) $\sigma_{\rm N} = (dI/dP) {\rm NEP} / \sqrt{2}$, where $\sqrt{2}$ converts from noise power spectra over $1/\sqrt{\rm Hz}$ to sensitivity in $\sqrt{\rm sec}$.}

\begin{figure}
    \centering
    \includegraphics[width = 0.9\columnwidth]{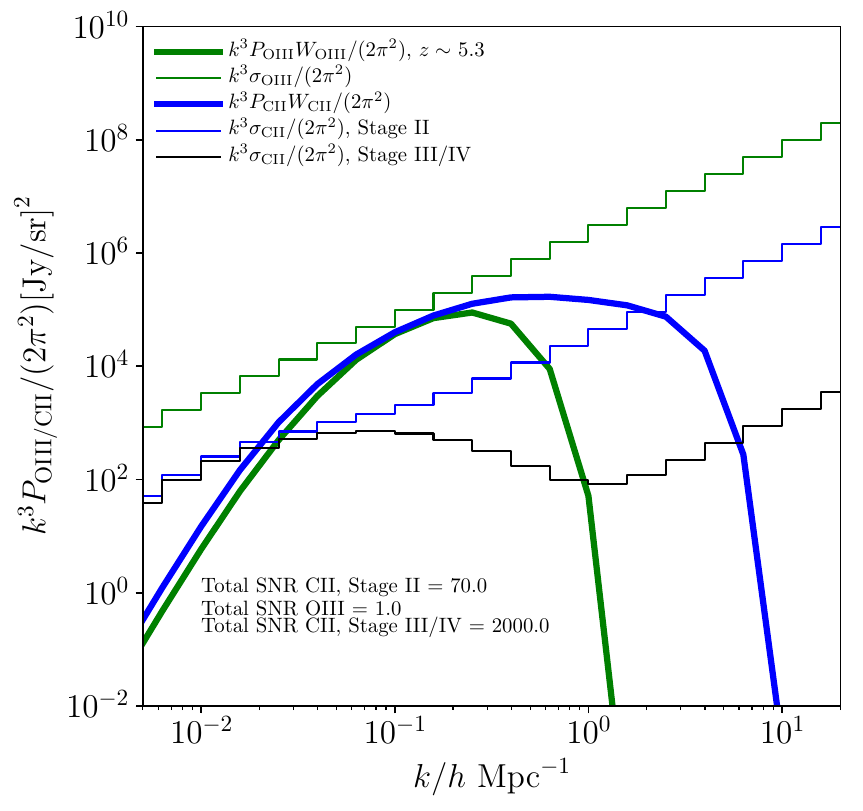} \includegraphics[width = 0.9\columnwidth]{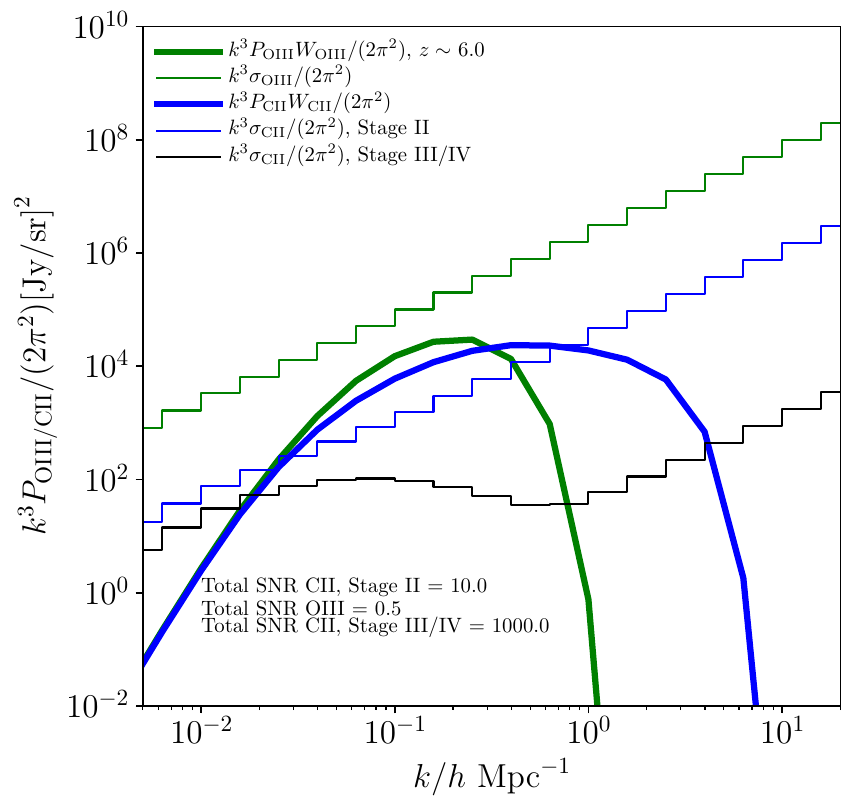}\\
    \includegraphics[width = 0.9\columnwidth]{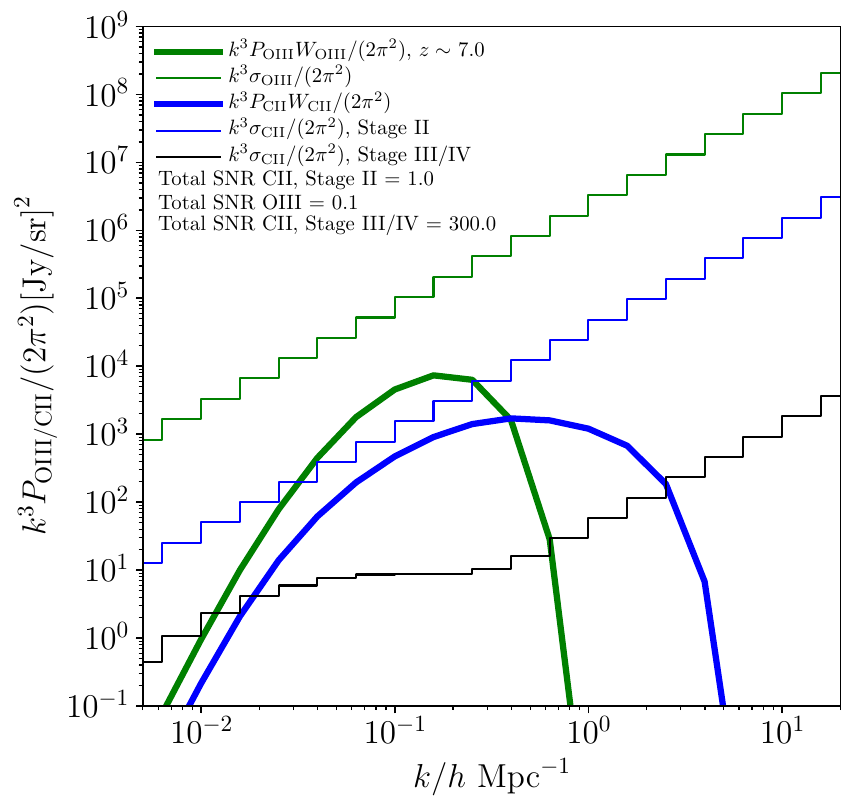}
    \caption{Auto-correlation signal power $k^3 P_{\rm CII}/(2 \pi^2)$ and $k^3 P_{\rm OIII}/(2 \pi^2)$ and the auto-correlation noise power, $k^3 \sigma_{\rm CII}/(2 \pi^2)$ and  $k^3 \sigma_{\rm OIII}/(2 \pi^2)$. { The plots are for improved versions of an EXCLAIM-like 72 h and FYST-like 2000 h Stage II and Stage III/IV surveys with parameters as defined in Table \ref{table:improved}, probing [OIII] and [CII] respectively at bands centred on $z \sim 5.3, 6$ and 7.} The total signal-to-noise ratios for both [CII] and [OIII] [calculated using \eq{snrauto}] are indicated on each plot.}
    \label{fig:autocorr}
\end{figure}

\subsection{Spatial and spectral resolution effects}
\label{sec:resolution}

\begin{figure}
    \centering
    \includegraphics[width = 0.9\columnwidth]{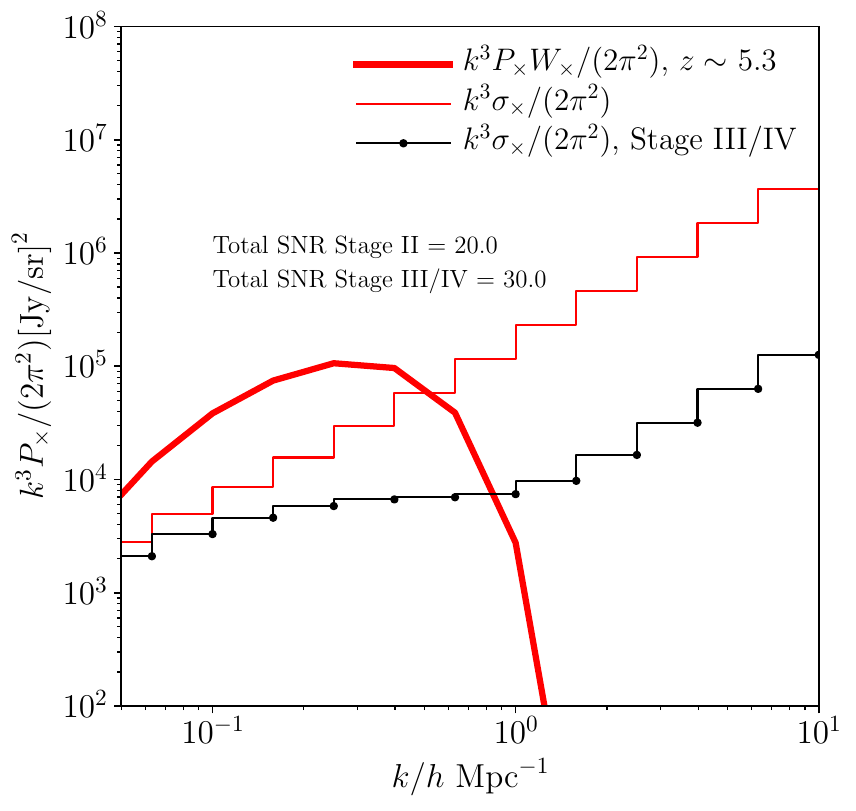} \includegraphics[width = 0.9\columnwidth]{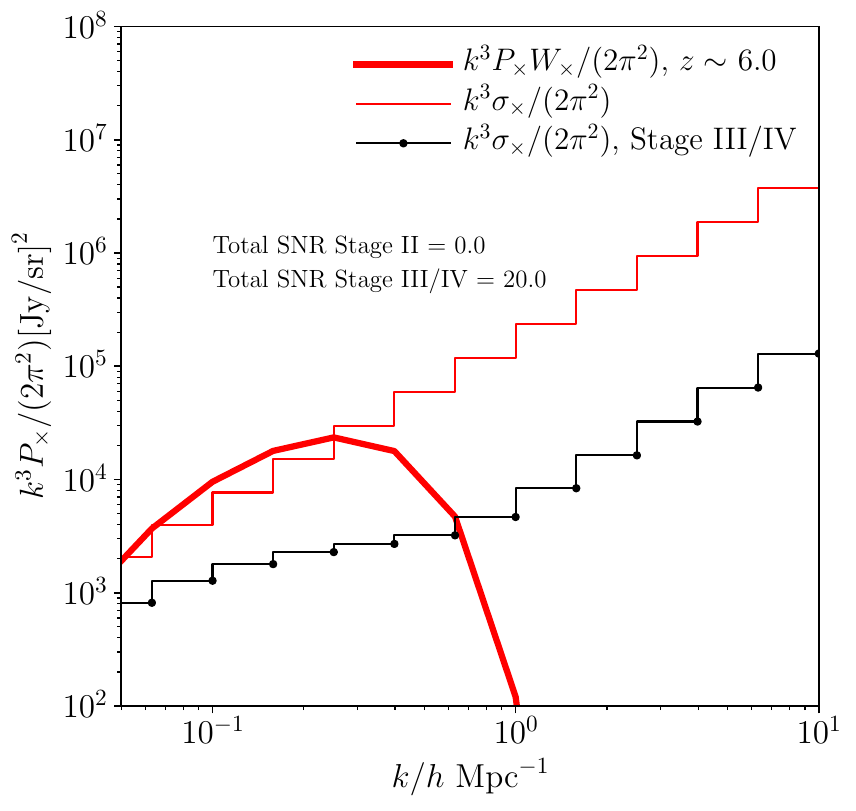}\\
    \includegraphics[width = 0.9\columnwidth]{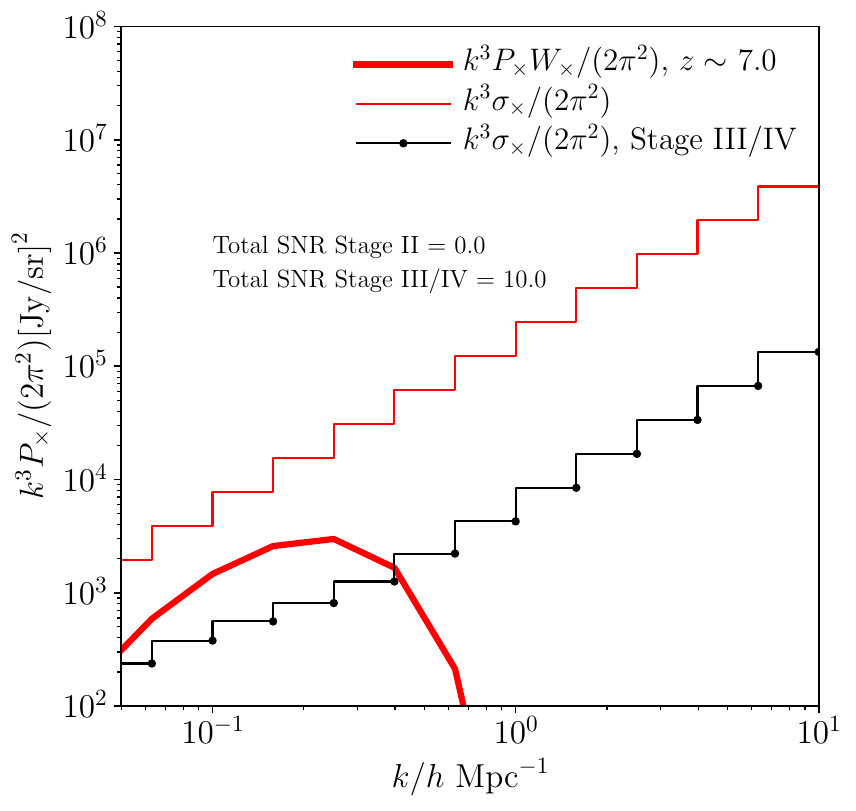}
    \caption{Cross-correlation signal (thick line) and standard deviation (thin steps) for the EXCLAIM-like and FYST-like configurations { with parameters as in Table \ref{table:improved}} at $z \sim 5.3, 6$ and 7. The total SNR (obtained by using \eq{snrcross}) is marked on each figure.}
    \label{fig:crosspowerexclaimfyst}
\end{figure}

The signal power spectra, \eq{powerspeccii} are attenuated by the effects of limited spatial and spectral resolutions in the survey. These effects are typically modelled through a weighting factor in $k$-space, defined by:

\begin{eqnarray}
 W_{\rm beam}(k) &=& e^{-k^2 \sigma_{\perp, j}^2} \int_0^1 e^{-k^2 \mu^2 (\sigma_{\parallel, ij}^2 - \sigma_{\perp, j}^2)} d \mu \nonumber \\
   \noindent &=& \frac{1}{k \sqrt{\sigma_{\parallel, ij}^2 - \sigma_{\perp, j}^2}} \frac{\sqrt{\pi}}{2} {\rm{Erf}} \left(k  \sqrt{\sigma_{\parallel, ij}^2 - \sigma_{\perp, j}^2} \right)  \nonumber \\
   &\times& \exp(-k^2 \sigma_{\perp, j}^2)
   \label{wij}
\end{eqnarray}
with $j \in \{\rm EXCLAIM, FYST\}$ depending on the parameters of the corresponding experiment. The second equality above assumes that $\sigma_{\parallel, ij} > \sigma_{\perp, j}$, which is valid for all cases considered here.
In the above expression, we have defined $
\mu = \cos \ \theta_k$, where $\theta_k$ is the polar angle in $k$-space. The quantities $\sigma_{\perp, j}$ and $\sigma_{\parallel, ij}$ are defined through \citep{li2015}:

\begin{equation}
    \sigma_{\perp, j} = R(z) \sigma_{\rm beam, j}
    \label{sigmaperp}
\end{equation}
where $R(z)$ is the comoving distance to redshift $z$, $\sigma_{\rm beam, j} = \theta_{\rm beam, ij}/\sqrt{8 \ln 2}$ as defined above, and 
\begin{equation}
    \sigma_{\parallel, ij} = \frac{c}{H(z)} \frac{(1+z)^2 \Delta \nu_j}{\nu_{\rm obs, i}}
    \label{sigmapar}
\end{equation}
where $H(z)$ is the Hubble parameter at redshift $z$, with the parameter $\Delta \nu_j$ given by that of the corresponding experiment and $\nu_{\rm obs,i}$ being the observed frequency of the corresponding transition ($i \in \{\rm [CII], [OIII]\}$).

\subsection{Finite volume effects}
\label{sec:volumeeffects}
{
To account for the finite volume of the survey, we introduce a window function in volume, $W_{\rm vol}$, defined through \citep[e.g.,][]{bernal2019}:

\begin{equation}
    \begin{split}
W_{\rm vol}(k,\mu) = & \left(1-\exp\left\lbrace -\left(\frac{k}{k^{\rm min}_\perp}\right)^2\left(1-\mu^2 \right) \right\rbrace \right)\times \\
& \times \left(1-\exp\left\lbrace -\left(\frac{k}{k^{\rm min}_\parallel}\right)^2\mu^2 \right\rbrace \right).
\end{split}
\label{eq:Wk_vol}
\end{equation}

{The window function above results from assuming an input survey sensitivity that is tapered like a Gaussian in configuration space. } In the above expression, $k^{\rm min}_{\perp} \equiv 2\pi/L_{\perp}$, and $k^{\rm min}_{\parallel} \equiv 2\pi/L_{\parallel}$,
where the volume of the survey is approximately given by $L_\perp^2 L_\parallel$ with $L_\perp$ and $L_\parallel$ being the maximum transverse and radial length scales probed by the survey.  These are defined by:
\begin{equation}
    L_{\parallel} = \frac{c}{H(z)} \frac{(1+z) B_{\nu, j}}{\nu_{\rm obs, i}}
\end{equation}
where $B_{\nu,j}$ is the bandwidth of the respective survey, and
\begin{equation}
    L_{\perp}^2 = \chi^2(z) \Omega_{A}
\end{equation}
where $\chi(z)$ is the comoving distance to redshift $z$, and $\Omega_A$ is the solid angle associated with the area of the survey.
Averaged over angular variable $\mu$, \eq{eq:Wk_vol} leads to
\begin{eqnarray}
   && W_{\rm vol}(k) = 1 - \exp(-k^2/k_{\perp}^2)  \int_0^1 d \mu \exp (k^2 \mu^2 /k_{\perp})^2 \nonumber \\
   &+& \exp(-k^2/k_{\perp}^2) \int_0^1 \exp - \left(k^2 \mu^2 /k_{\parallel}^2 - k^2 \mu^2/k_{\perp}^2\right) d \mu \nonumber \\
    &&  - \int_0^1d \mu  \exp  (- k^2 \mu^2 /k_{\parallel}^2) \nonumber \\
    &\approx & 1 - \frac{\sqrt{ \pi} k_{\parallel}}{2 k}{\rm{Erf}} \left(k / k_{\parallel} \right)
\end{eqnarray}
(we have suppressed the  superscripts  "min"  in $k_{\perp}$ and $k_{\parallel}$ in the above equation for simplicity)
and thus to a similar expression  as in \eq{wij} above for  $k^{\rm min}_{\parallel}  \gg k^{\rm min}_{\perp}$, which holds for all the cases under consideration. [Note that the last equation above is an excellent approximation in the regime of interest.]

}
\begin{table*}
\begin{center}
    \begin{tabular}{ | c | c | c | c | c | c | c | c| c|  p{0.5cm} |}
    \hline
    Configuration &  $D_{\rm dish}$ (m.) & $\Delta \nu$ (MHz) &  $N_{\rm spec, 
eff} $ & $S_{\rm A}$ (sq. deg.) & $\sigma_{\rm N}$ 
(Jy s$^{1/2}$ / sr) & $B_{\nu}$ (GHz) & $t_{\rm obs}$ (h.) & Survey Bandwidth \\ \hline
   
    Design & 3 & 300 & 50 & 16 & $1.6 \times 10^5$  & 100 & 4000 &  250-900 GHz \\
\hline
    \end{tabular}
\end{center}
\caption{`Design' parameters for an [OIII] $\times$ [CII] intensity mapping survey as described in the main text.}
 \label{table:designexpt}
\end{table*}

The signal power in each case is modulated through $W_{\rm ij}(k) = W_{\rm beam} (k) W_{\rm vol}(k)$.

From the signal and noise power spectra, \eq{noiseciiauto} and \eq{powerspeccii} we can construct the variance of the noise, which is  defined as:
\begin{equation}
    \text{var}_{\rm ij} = \frac{(P_{\rm i} W_{\rm ij}(k) + P_{\rm N, ij})^2}{{N_{\rm modes, ij}}}
    \label{varianceauto}
\end{equation}
where the number of modes, $N_{\rm modes, i}$ is defined for each survey as:
\begin{equation}
    N_{\rm modes, ij} = 2 \pi k^2 \Delta k \frac{V_{\rm surv, ij}}{(2 \pi)^3} 
\end{equation}
with the volume of the survey, $V_{\rm surv, ij}$ given by:
\begin{eqnarray}
V_{\rm surv, ij} &=& 3.7 \times 10^7  {\rm (cMpc}/h)^3 \left (\frac{\lambda_i}{158  \ 
\mu {\rm m}} \right) \left(\frac{1 +z}{8} \right)^{1/2}  \nonumber \\
&& \left(\frac{S_{\rm{A, j}}}{16 {\rm deg}^2} \right) \left(\frac{B_{\nu, j}}{20 
{\rm GHz}} \right)
\label{vsurveyauto}
\end{eqnarray}
In the above expression, $\Delta k$ denotes the bin width in $k$-space (we use logarithmically equispaced $k$-bins with $\Delta \log_{10} k = 0.2$ for all the cases considered here). { Note that the above  expression makes the implicit assumption of Gaussianity for both the signal and noise terms, as is frequently done in the literature \citep{dumitru2018, hpcii2019, bernal2019, breysse2021}. Any additional effects due to non-Gaussianity are likely to be subdominant to the model uncertainties, especially in the detector noise dominated limit. Moreover, the suppression of the beam on small scales (\eq{wij}) ensures that the information obtainable from the shot noise regime -- where such effects are expected to be important -- is subdominant to that coming from larger scales. }

We plot the signal power spectra with angular and spectral resolution effects applied for [CII] and [OIII] as $k^3 P_{\rm CII} W_{\rm CII, FYST}(k)/2 \pi^2)$ and $k^3 P_{\rm OIII} W_{\rm OIII, EXCLAIM}(k)/2 \pi^2)$ respectively by the solid blue, black and green lines in Fig. \ref{fig:autocorr}. The three panels are centred at $z \sim 5.3, 6$ and 7 respectively, corresponding to the redshift ranges ${z_{\rm min}- z_{\rm max}} = \{4.9 - 5.7, 5.5 - 6.5, 6.3 - 7.8\}$.

Overplotted are the noise power spectra $k^3 \sigma_{\rm CII, FYST}/(2 \pi^2)$ and $k^3 \sigma_{\rm OIII, EXCLAIM}/(2 \pi^2)$ at $z \sim 5.3, 6$ and 7 shown by the thin steps, where $\sigma_{\rm ij} = (\rm var_{\rm ij})^{1/2}$.

The total signal-to-noise ratio, given by
\begin{equation}
    {\rm{SNR}} = \left(\sum_{k} \frac{P_{\rm i}^2 W^2_{\rm ij} (k)}{{\rm{var}}_{\rm ij}} \right)^{1/2}
\label{snrauto}
\end{equation}
(where the sum is over all the $k$-modes under consideration) is indicated on each plot for both [OIII] and [CII]. The autopower forecasts indicate that the [CII] power spectrum detection is possible to a high significance { at all the redshifts under consideration for the Stage III/IV survey, and out to $z \sim 6$ for the Stage II configuration}. For [OIII], an auto-power detection with the improved EXCLAIM-like configuration may be possible at $z \sim 5.3$.

\section{Cross-correlating a ground-based [CII] and a balloon-based [OIII] survey}
\label{sec:crosscorr}

We now consider a cross-correlation of [OIII] 88 $\mu$m  and [CII] 158 $\mu$m as observed together with improved versions of the EXCLAIM-like and FYST-like configurations above.  Given the power spectrum in \eq{powerspeccii}, we can calculate the cross-correlation signal (at a given redshift) as \footnote{Note that we are neglecting the difference between the averaging over the product of the luminosities in \eq{CIIspint} and that of the derived power spectra. Incorporating this difference can be shown to lead to a shot-noise-like contribution to the cross-power \citep[see, e.g.,][]{liu2020}, but this is unimportant for our present analysis since our signal is mostly in the clustering regime.}:
\begin{equation}
    P_{\times}(k) = (P_{\rm CII}(k) P_{\rm OIII}(k))^{1/2}
    \label{crosspower}
\end{equation}
{ The above equation assumes that the two fields are highly correlated ($r{\approx}1$). This assumption is expected to be valid on large scales where the stochasticity is negligible, and linear biasing is a good approximation. On small scales, it may break down due to the two lines tracing different parts of the interstellar medium \citep[see, e.g.][]{breysse2021}, with [OIII] coming from the HII regions and [CII] from the photon-dominated regions (PDRs), and furthermore, the two lines may originate partly from different host dark matter halo populations. However, given the suppression of the signal due to beam effects on small scales, the information content from this regime is subdominant to that from larger scales.}

We consider this configuration observing at $z \sim 5.3, 6$ and 7, with fiducial observing times of 2000 h and 72 h for the FYST-like [CII] and EXCLAIM-like [OIII] surveys respectively. We assume further that the two surveys overlap entirely, which means that the EXCLAIM-like survey focuses its entire observing time\footnote{Note that this differs from the base EXCLAIM configuration which will spend its observing time of 8 h on a region of 400 deg$^2$.}  of 72 h on the Stage II [CII] survey's patch of sky, which covers 100 deg$^2$. 

To calculate the variance of the cross power spectrum, we need the number of modes in the cross-correlation survey, defined through:
\begin{equation}
N_{\rm modes, \times} = 2 \pi k^2 \Delta k \frac{V_{\rm surv, \times}}{(2 \pi)^3}
\label{nmodes}
\end{equation}
with $\Delta k$ being the bin width in $k$-space (using equally-spaced logarithmic $k$-bins with $\Delta \log_{10} k 
= 0.2$), and the survey volume \citep{gong2012, dumitru2018} given by:
\begin{eqnarray}
V_{\rm surv, \times} &=& 3.7 \times 10^7  {\rm (cMpc}/h)^3 \left (\frac{\lambda}{158  \ 
\mu {\rm m}} \right) \left(\frac{1 +z}{8} \right)^{1/2}  \nonumber \\
&& \left(\frac{S_{\rm{A, \times}}}{16 {\rm deg}^2} \right) \left(\frac{B_{\nu}}{20 
{\rm GHz}} \right)
\label{vsurvey}
\end{eqnarray}
To calculate $V_{\rm surv, \times}$, we use $\lambda = 158 \mu {\rm m}$, $S_{\rm A, \times} = 100$ deg$^2$, and $B_{\nu} = 40$ GHz (corresponding to  the Stage II [CII] configuration).

The variance in the cross-correlation is then given by:
\begin{equation}
    {\rm var}_{\times} = \frac{(P_{\rm CII} + P_{\rm N(CII), FYST}) (P_{\rm OIII} + P_{\rm N(OIII), EXCLAIM}) + P_{\times}^2}{2 N_{\rm modes, \times}} 
\label{varcii}
\end{equation}
with $P_{\rm N(CII), FYST}$ (for both FYST-like configurations) and $P_{\rm N(OIII), EXCLAIM}$ following \eq{noiseciiauto}.
If we account for the finite spatial and spectral resolution, { as well as volume effects} (see Secs. \ref{sec:resolution} and \ref{sec:volumeeffects}), the above expression gets modified to:
\begin{eqnarray}
    {\rm var}_{\times}(k) &=& \left((P_{\rm CII}W_{\rm CII, FYST}(k) + P_{\rm N(CII), FYST}) \right. \nonumber\\
   & & \left. (P_{\rm OIII}W_{\rm OIII, EXCLAIM}(k) + P_{\rm N(OIII), EXCLAIM}) \right. \nonumber\\
    &+& \left. P_{\times}^2 W_{\times}^2(k) \right)/2 N_{\rm modes, \times}
\label{varcii}
\end{eqnarray}
where we have defined 
\begin{equation}
    W_{\times}(k) = (W_{\rm CII, FYST}(k) W_{\rm OIII, EXCLAIM} (k))^{1/2}
\end{equation}
analogously to $P_{\times} (k)$.

\begin{figure}
    \centering
    \includegraphics[width = 0.9\columnwidth]{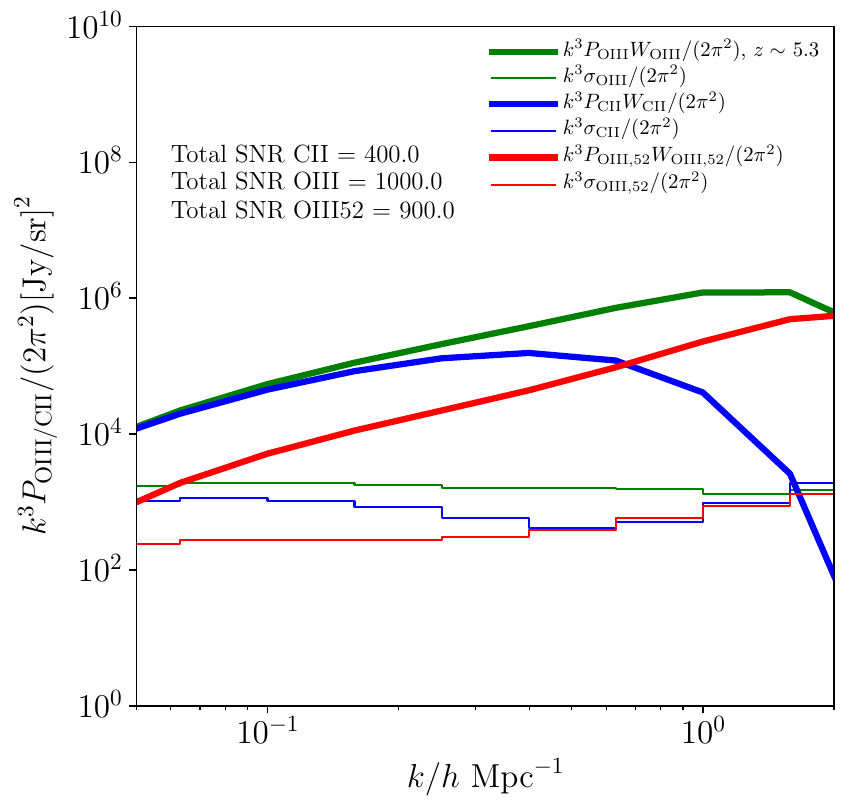} \includegraphics[width = 0.9\columnwidth]{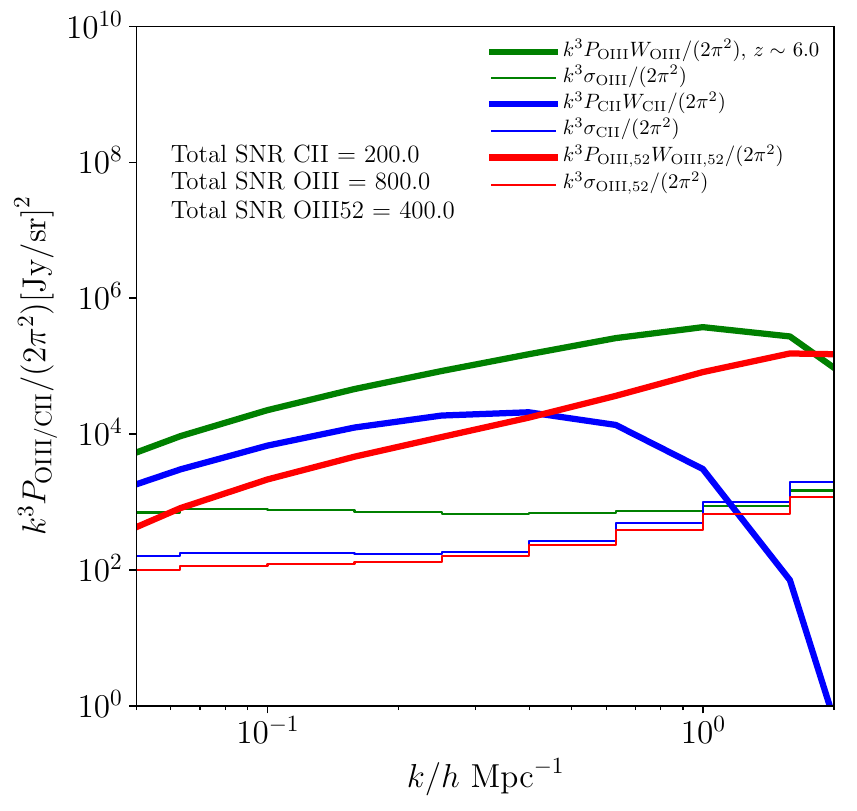}\\
    \includegraphics[width = 0.9\columnwidth]{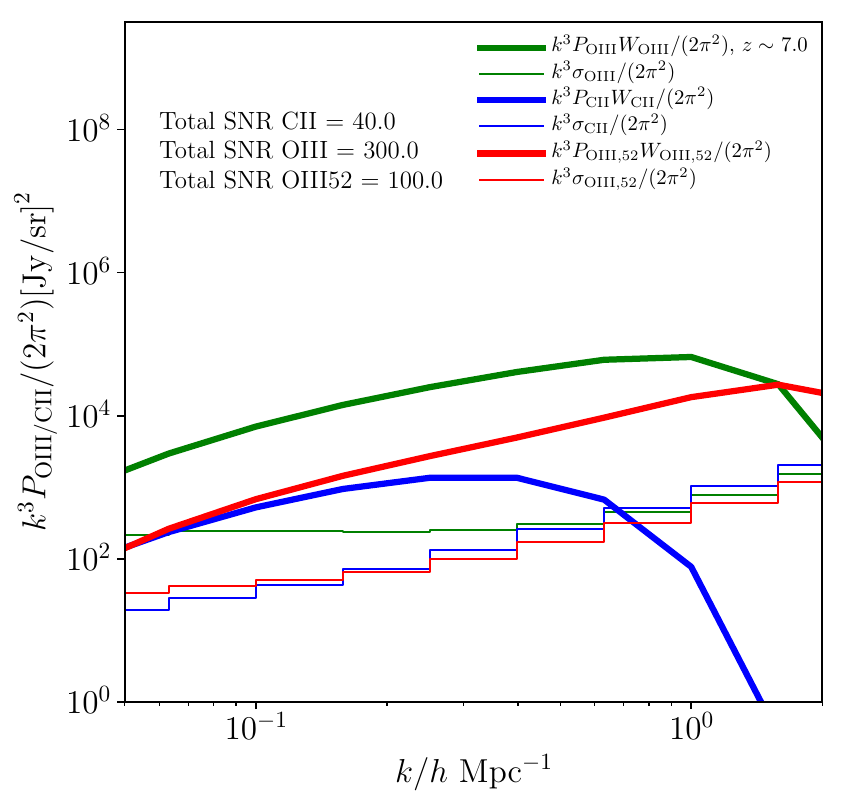}
    \caption{Auto-correlation signal power $k^3 P_{\rm CII}/(2 \pi^2)$, $k^3 P_{\rm OIII}/(2 \pi^2)$ and $k^3 P_{\rm OIII, 52}/(2 \pi^2)$ and the auto-correlation noise power, $k^3 \sigma_{\rm CII}/(2 \pi^2), k^3 \sigma_{\rm OIII}/(2 \pi^2)$ and  $k^3 \sigma_{\rm OIII, 52}/(2 \pi^2)$. The plots are for the design configuration at bands centred on $z \sim 5.3, 6$ and 7 respectively. The total SNR (obtained from \eq{snrauto}) is indicated on each plot.}
    \label{fig:autocorrdesign}
\end{figure}

Fig. \ref{fig:crosspowerexclaimfyst} plots the cross correlation signal ($k^3 P_{\times} W_{\times}/2 \pi^2$) and its standard deviation ($k^3 \sigma_{\times}/2 \pi^2$) for the improved EXCLAIM-FYST like experiment combinations for bands centred at $z \sim 5.3, 6$ and 7 [with $\sigma_{\times} = (\rm var_{\times})^{1/2}$]. Also mentioned on each plot are the expected SNR for each case, calculated as:
\begin{equation}
    {\rm{SNR}} = \left(\sum_k \frac{P_{\times}^2(k) W_{\times}^2(k)}{{\rm{var}}_{\times}(k)}\right)^{1/2} \, ,
    \label{snrcross}
\end{equation}
where the sum is over the $k$-modes under consideration. 
It can be seen that significant detections are possible at all redshifts ($z \sim 5.3, 6$ and 7) { with the FYST-like Stage III/IV configuration, and at $z \sim 5.3$ with the Stage II configuration}.

\section{An `ideal' [OIII] and [CII] intensity mapping survey?}
\label{sec:design}

We now consider a designed [CII] - [OIII] cross-correlation survey, with the parameters as provided in Table \ref{table:designexpt}. This survey is tailored to optimize on both the EXCLAIM-like and FYST-like surveys above so that the signal-to-noise ratio can be maximized. We envision a future space-based mission for this configuration, giving background-limited predictions { motivated by Fig. \ref{fig:nei_scenarios}}. This experiment may also be particularly well-suited to target other lines, notably the [OIII] 52 $\mu$m line, another fine-structure line tracing the same ion of [OIII]. 

With this in mind, for the `design' survey, in addition to [OIII] 88 $\mu$m and [CII] 158 $\mu$m, we also explore forecasts with the [OIII] 52 $\mu$m line.  The main difference between this line and the [OIII] 88 $\mu$m line lies in their critical densities (above which collisional de-excitations outpace spontaneous decays). The ratio of the two line luminosities depends primarily on the electron density of the emitting regions, varying between $L_{\rm [OIII]52}/L_{\rm [OIII]88} \sim 0.55$ at low densities to $\sim 10$ at high densities \citep[e.g.,][]{yang2020}. For simplicity, we consider the lower limit in our present forecasts for [OIII] 52 $\mu$m, $L_{\rm [OIII]52}/L_{\rm [OIII]88} \sim 0.55$, corresponding to the most pessimistic low-density case. 

\begin{figure}
    \centering
    \includegraphics[width = 0.9 \columnwidth]{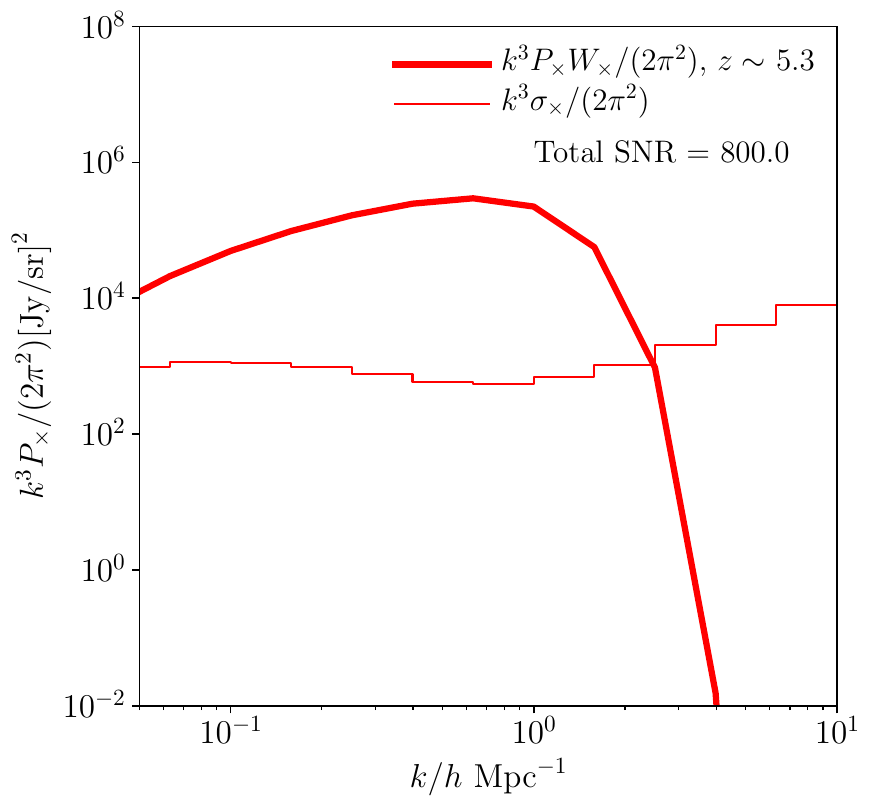} \includegraphics[width = 0.9 \columnwidth]{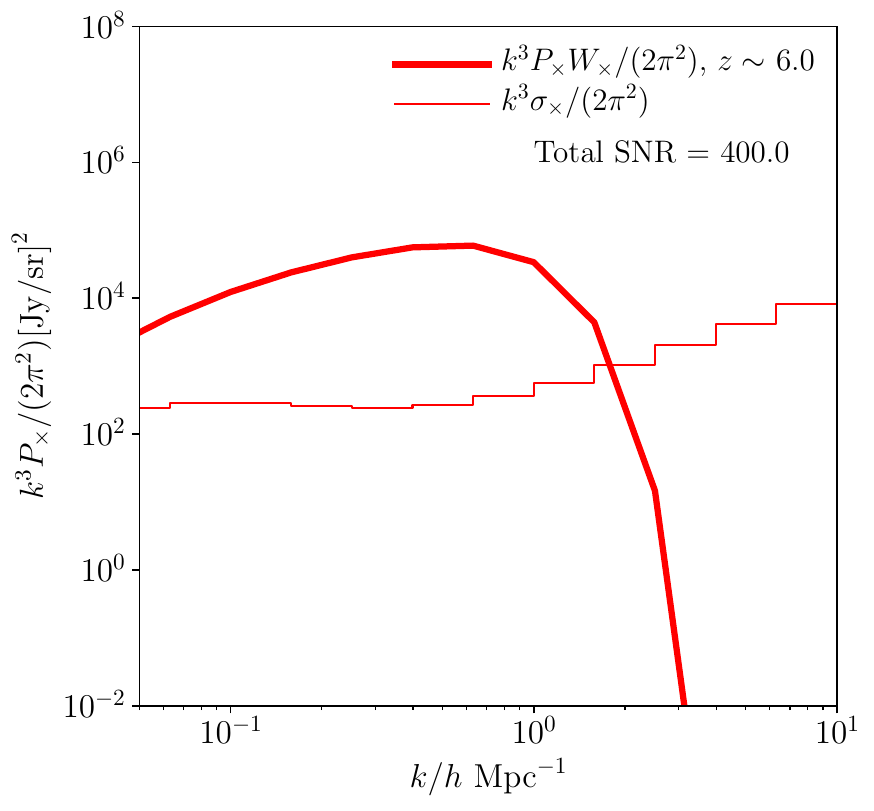} \includegraphics[width = 0.9 \columnwidth]  {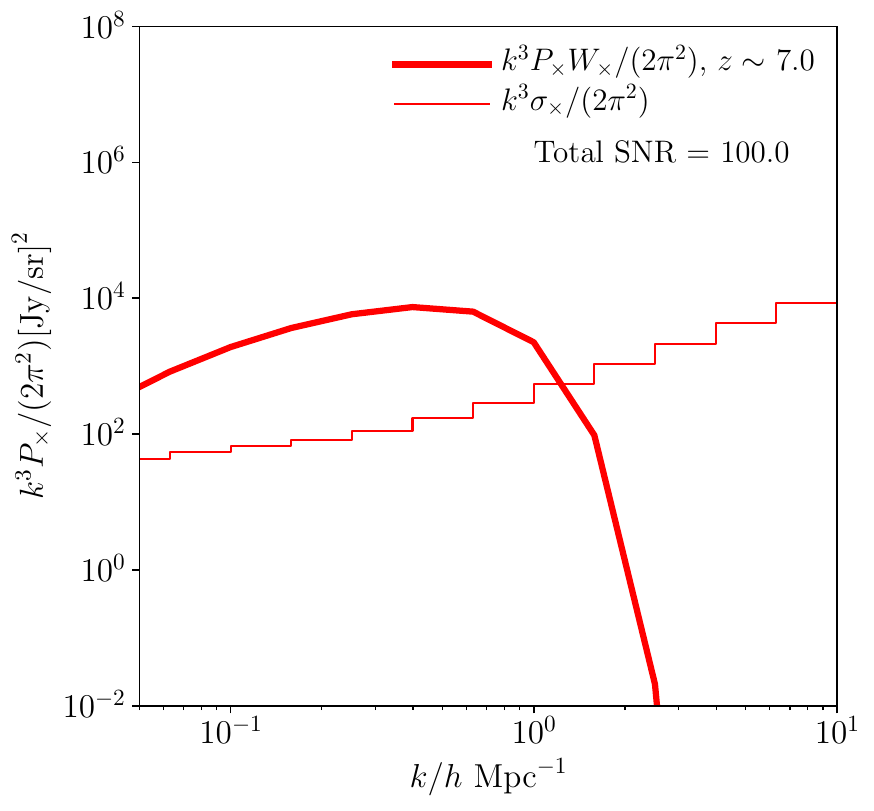}
    \caption{Cross-correlation signal (thick line) and standard deviation (thin steps) for a `design' experiment combination (with parameters specified in Table \ref{table:designexpt}) at $z \sim 5.3, 6$ and 7 (from top to bottom) observing [CII] 158 $\mu$m and [OIII] 88 $\mu$m. The total SNR (obtained from \eq{snrcross}) is also shown in all cases.}
    \label{fig:crosspowerdes}
\end{figure}

\begin{figure}
    \centering
    \includegraphics[width = 0.9 \columnwidth]{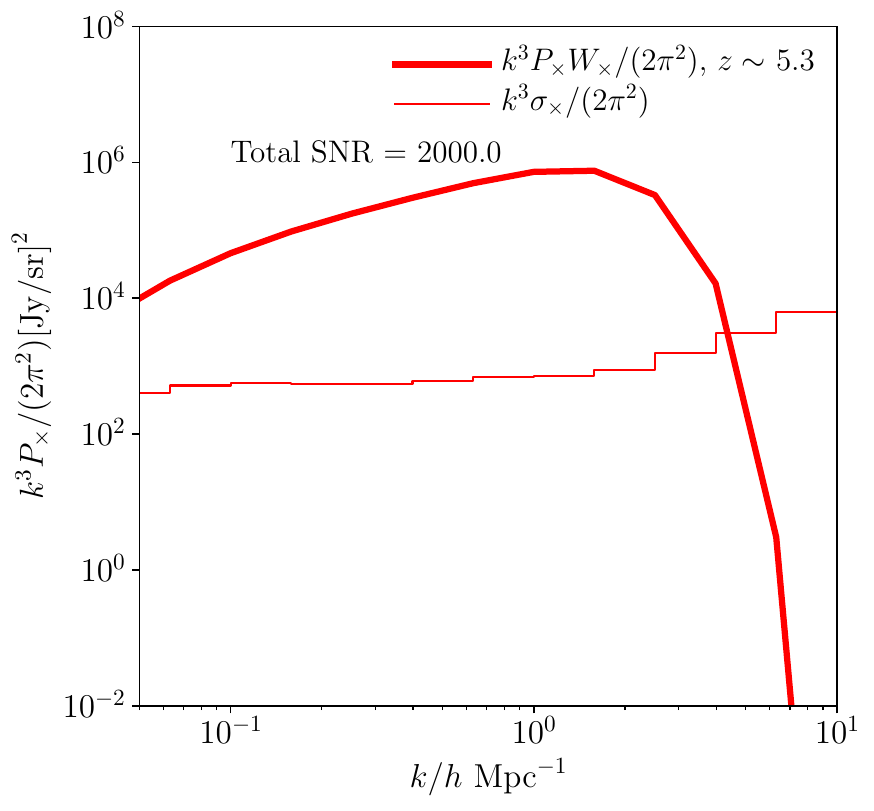} \includegraphics[width = 0.9 \columnwidth]{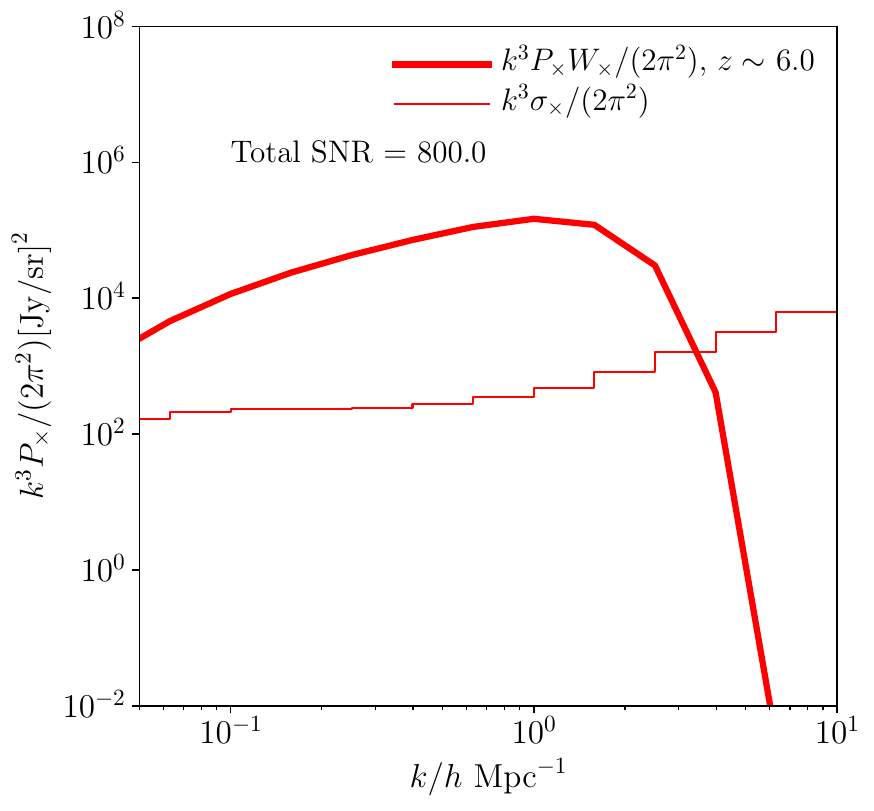} \includegraphics[width = 0.9 \columnwidth]  {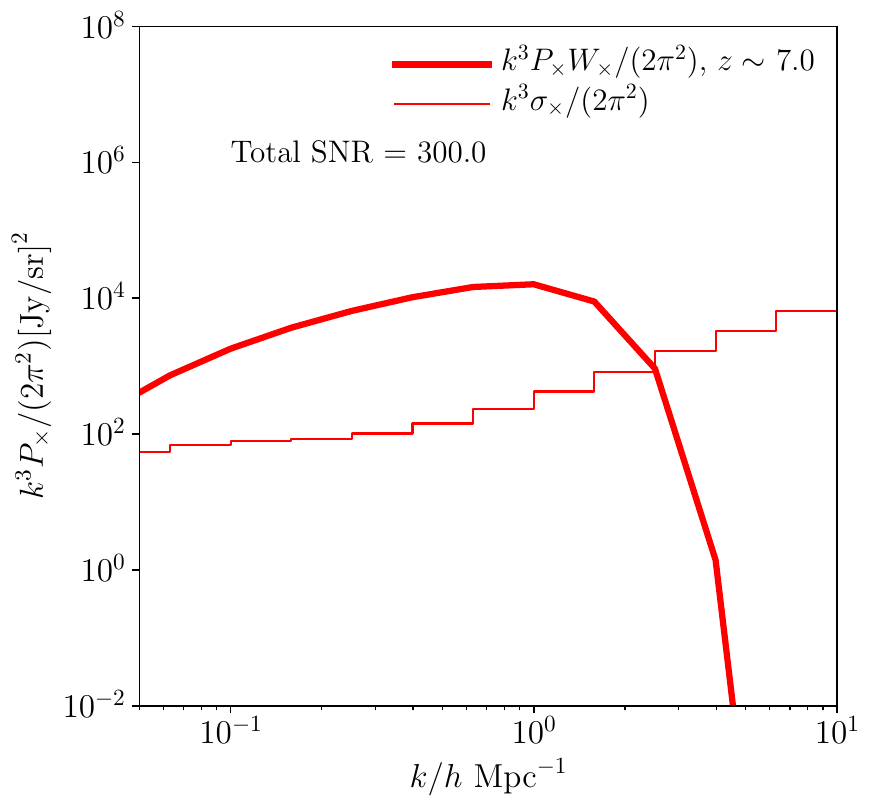}
    \caption{Cross-correlation signal (thick line) and standard deviation (thin steps) for a `design' experiment combination (with parameters specified in Table \ref{table:designexpt}) at $z \sim 5.3, 6$ and 7 (from top to bottom) observing [OIII] 52 $\mu$m and [OIII] 88 $\mu$m. The total SNR (obtained from \eq{crosspower} - \eq{snrcross}, with [CII] replaced by [OIII] 52 $\mu$m) is also shown in all cases.}
    \label{fig:crosspoweroiiilines}
\end{figure}

\subsection{Auto-power spectra}
Here, we estimate the expected auto-power spectra and noise for the `design' survey configuration, in [CII] and both lines of [OIII], 88 $\mu$m and 52 $\mu$m at $z \sim 5.3, 6$ and 7. We adopt the same formalism as in Sec. \ref{sec:formalism} to model the signal power, with the only differences (for [OIII] 52 $\mu$m) being the wavelength and the luminosity of the line (scaled down from that of the [OIII] 88 $\mu$m by a factor 0.55). Figs. \ref{fig:autocorrdesign} show these forecasts for each of the three lines above observed with the design survey. The thin steps on each plot show the noise of the configuration, $k^3 \sigma_{\rm CII}/(2 \pi^2)$, $k^3 \sigma_{\rm OIII}/(2 \pi^2)$ and $k^3 \sigma_{\rm OIII, 52}/(2 \pi^2)$, where $\sigma_{\rm ij} = (\rm var_{\rm ij})^{1/2}$ with $\rm var_{\rm ij}$ given by \eq{varianceauto}.
The total signal-to-noise ratio calculated by using \eq{snrauto} is indicated on each plot. It can be seen that there is a significant improvement in the signal-to-noise expected for each of the lines in this scenario, as compared to the cases with the EXCLAIM-like and FYST-like configurations investigated previously in Fig. \ref{fig:autocorr}.

\subsection{Cross-power spectra}

We now consider two cross-correlation configurations with the above `design' survey: (a) between [CII] 158 $\mu$m and [OIII] 88 $\mu$m, and (b) between [OIII] 52 $\mu$m and [OIII] 88 $\mu$m. 
The first configuration, identical to the survey we considered in Sec. \ref{sec:crosscorr} with the improved EXCLAIM-like and FYST-like [CII] parameters, is plotted in Figs. \ref{fig:crosspowerdes} (with the same notation as before). The total SNR improves significantly as compared to the previous case in Fig. \ref{fig:crosspowerexclaimfyst}. Even at $z \sim 7$, one can achieve a signal-to-noise ratio of $\sim 100$ when summed across all $k$-modes.

The power spectra and noise for the second configuration, [OIII] 52 $\mu$m cross-correlated with [OIII] 88 $\mu$m at $z \sim 5.3, 6$ and 7 are plotted in Figs. \ref{fig:crosspoweroiiilines}. The total SNR, (calculated from \eq{crosspower} - \eq{snrcross}, with [CII] replaced by [OIII] 52 $\mu$m) is also indicated on each plot. We find that this configuration also  leads to a large improvement in the SNR at all redshifts. This cross-correlation scenario is especially useful to reject line interlopers while being cleaner to interpret than the [OIII] 52 $\mu$m $\times$ [CII] 158 $\mu$m case considered above, and would thus be an ideal target for a future space-based mission.

\section{Conclusions}

This paper has explored the potential of [OIII] as a strong tracer of large-scale structure relevant to intensity mapping at $z > 5$ targeted by upcoming surveys. We have focused on the cross-correlation between the [OIII] and [CII] tracers, with fiducial survey configurations { improving upon} the upcoming FYST (in [CII]) and EXCLAIM (for [OIII]) experimental configurations. We found that enhanced, next-generation versions of these ground- and balloon-based missions may enable significant cross-correlation detections at the $\sim 10 - {30} \sigma$ confidence level for $z > 5$. 

We also explored the potential for detecting [OIII] with a `design', or near-ideal space-based survey which is background-limited. Particularly, such an experiment is extremely well suited to target other lines of interest at submillimetre frequencies. We chose the fiducial example of the [OIII] 52 $\mu$m line, which is extremely useful as an additional probe of the same regions as the [OIII] 88 $\mu$m line, which also traces the same ion. Since the main difference between these two lines lies only in their critical densities, the [OIII] 52 $\mu m$ - [OIII] 88 $\mu$ m cross-correlation would reject line interlopers, but be cleaner to interpret than the [OIII] 88 $\mu$m - [CII] 158 $\mu$m. { The presence of two lines also mitigates the challenge of possible contamination from [CII] at lower redshifts (in the case of EXCLAIM, $z \sim 2.5 - 3.5$) to a large extent.} Furthermore, the ratio of the [OIII] 88 to [OIII] 52 $\mu$m strengths contains information about the typical electron densities in the HII regions of these early galaxy populations. 
Both these configurations were found to lead to large improvements in the signal-to-noise ratio as compared to those with present facilities, demonstrating the effectiveness of the design experiment and its preferred targets for auto- and cross-correlation studies. 

{

Although we defer a more detailed treatment regarding the impact of foreground contamination to possible future work, a few comments are appropriate here. First, we can consider spectrally smooth continuum emission from e.g. Galactic dust and the cosmic infrared background. This can be mitigated by avoiding low $k_\parallel$ modes in the observed data cubes: although this approach sacrifices measuring the signal for such modes, current work suggests that this only slightly degrades the expected SNR for CII line-intensity mapping measurements \citep{dizgah2018}. The case of OIII should be similar. The second is line-interloper contamination which may present a larger concern for CII and OIII measurements. In the case of CII surveys, the dominant interloper lines are expected to be CO rotational transitions at a range of lower redshifts. In the $z \sim 6-8$ regime (primarily accessible to FYST), the maps may be contaminated by interloping CO rotational line emission ($3 < J_{\rm upp} < 6$)  from galaxies at $0 < z < 2$. 

The OIII 88 and 52 micron observations will be contaminated both by CO fluctuations as well as CII emission from gas at lower redshift.  In the absence of mitigation measures, estimates of the auto power spectra in each line would be biased by the line interloper fluctuations. On the other hand, as alluded to earlier, cross power spectra between CII and OIII, as well as that between the 88 and 52 micron lines, will be mostly unbiased on average. This is because the interloping lines for e.g. CII and OIII are -- for the most part -- widely enough separated in redshift to be statistically independent and uncorrelated. A caveat here is that in some cases, two different CO transitions interloping in CII and OIII, respectively, may lie relatively nearby in redshift. In future work, it may be interesting to consider whether this could lead to a non-negligible average bias in cross-spectrum estimates. In any case, the interloper fluctuations, if left unchecked, would still contribute to the variance of each cross power spectrum estimate.

Although a number of interloper mitigation strategies are currently being considered, one approach is to use traditional near-infrared galaxy surveys to help identify prominent CO emitters \citep[e.g.,][]{sun2018}. The general idea here is that the near-infrared emission will be correlated with the CO contamination and so one can mask voxels in the line-intensity map around infrared-luminous galaxies and thereby reduce the CO contamination. Ideally, the near-infrared observations should cover the entire sky area of the line-intensity mapping survey with relatively precise photometric redshift estimates \citep{sun2018}. Quantitatively, those authors find that this targeted masking approach can reduce the CO fluctuation level to less than 10\% of the CII power (at a representative wavenumber of $k=0.1 h {\rm Mpc}^{-1}$), while only removing 8\% of the survey volume. If it is feasible to apply this targeted masking across the large fields envisioned in our study with a similar level of success, we would expect little impact on our SNR forecasts.

}

Our results serve as a useful benchmark for optimising present and future surveys targeting a holistic picture of the interstellar medium, especially around the epoch of reionization. Since the intensity mapping surveys considered here provide constraints over the global properties of galaxies over large cosmic volumes, cross-correlating them with deep, targeted galaxy surveys conducted using ALMA and its successors like e.g., the Atacama Large Aperture Submillimeter Telescope \citep[AtLAST;][]{klaassen2019} as well as the next generation Very Large Array (ngVLA)\footnote{https://ngvla.nrao.edu/} would lead to exciting insights on the nature of the first ionizing sources. Complementary information from deep galaxy detections has been shown to significantly aid the interpretation of power spectrum results from intensity mapping \citep{pavesi2018, uzgil2019, gonzalez2019} by providing exquisite constraints on both the clustering and shot-noise components of the power spectrum in the future.

{ While we do not discuss the prospects for parameter estimation in detail in the present manuscript, a few points are worth mentioning in this regard. Firstly, the evolution of the auto and cross--correlation maps in [CII] and [OIII] would be a valuable dataset with which to address the origin of the `[CII] deficit' seen in high-redshift galaxies, and as a complement to targeted galaxy observations at these epochs. [OIII] luminosities at $z \sim 6-9$ are sensitive to the average metallicity, ionizing spectrum and properties of the HII regions associated with the reionization-era galaxy population \citep{yang2020}.} It is also of interest to explore the feasibility of cross-correlating three separate lines, differing in rest wavelength by a factor of $\sim 3$, given the wide frequency coverage required. Possible targets include the [OI] 63 $\mu$m and [NII] 122 $\mu$m lines, within the scope of future space-based missions.

\section*{Acknowledgements}
{We thank the referee for a detailed and helpful report.}
HP acknowledges support from the Swiss National Science Foundation via Ambizione Grant PZ00P2\_179934. AL acknowledges support through NASA ATP grant 80NSSC20K0497.

\section*{Data availability}
No new data were generated or analysed in support of this research. The software underlying this article will be shared on reasonable request to the corresponding author.

\bibliographystyle{mnras}
\bibliography{mybib}

\label{lastpage}
\end{document}